\documentclass[aps,prl,floats,twocolumn,showpacs,superscriptaddress,reprint]{revtex4}
\usepackage{graphicx,epsfig}
\usepackage{graphics,dcolumn,bm,epic, eepic,fleqn,float}
\usepackage{amssymb,amsmath,amsfonts,multirow,rotate,color}
\usepackage{soul}
\usepackage{wasysym}
\usepackage{xcolor}

\newcommand{\avg}[1]{\langle #1 \rangle}

\newcommand{\lay}[1]{^{[#1]}}

\usepackage{color}

\definecolor{darkblue}{rgb}{0,0,0.555}
\usepackage[colorlinks=true,
            linkcolor=black,
            urlcolor=blue,
            citecolor=blue,
            bookmarks=true,
            pdfborder={0 0 0}]{hyperref}

\def\Erdos{Erd\"os}

\begin{document}

\title{Pareto optimality in multilayer network growth}

\author{Andrea Santoro}
\affiliation{School of Mathematical Sciences, Queen Mary University of
  London, Mile End Road, E1 4NS, London (UK)}
\affiliation{Scuola Superiore di Catania, Universit\`a di Catania, Via Valdisavoia 9, 95125, Catania (Italy)}
\author{Vito Latora}
\affiliation{School of Mathematical Sciences, Queen Mary University of
  London, Mile End Road, E1 4NS, London (UK)}
\affiliation{Dipartimento di Fisica ed Astronomia, Universit\`a di Catania and
  INFN, I-95123 Catania, (Italy)}
\author{Giuseppe Nicosia}
\affiliation{Dipartimento di Matematica ed Informatica, Universit\`a di Catania, V.le A. Doria 6, 95125, Catania (Italy)}
\affiliation{Department of Computer Science, University of Reading, Whiteknights, RG6 6AF Reading, (UK)}
\author{Vincenzo Nicosia}
\affiliation{School of Mathematical Sciences, Queen Mary University of
  London, Mile End Road, E1 4NS, London (UK)}


\date{\today}

\begin{abstract}
  We model the formation of multi-layer transportation networks as a
  multi-objective optimization process, where service providers
  compete for passengers, and the creation of routes is determined by
  a multi-objective cost function encoding a trade-off between
  efficiency and competition. The resulting model reproduces well 
  real-world systems as diverse as airplane, train and bus networks, thus
  suggesting that  such systems are indeed compatible with the
  proposed local optimization mechanisms. In the specific
    case of airline transportation systems, we show that the networks
  of routes operated by each company are placed very close to the
  theoretical Pareto front in the efficiency-competition plane,
  and that most of the largest carriers of a continent belong
    to the corresponding Pareto front. Our results shed light on the
  fundamental role played by multi-objective
  optimization principles in shaping the structure of large-scale
 multilayer transportation systems, and provide novel insights  to service 
 providers on the strategies for the smart selection of novel routes.
\end{abstract}

 \maketitle

The interactions among the basic units of many natural and man-made
systems, including living organisms, ecosystems, societies, cities,
and transportation systems are well described by complex
networks~\cite{BA_2002,NewmanSIAM_2003,Boccaletti_2006,Newman_2010,lnr_2017}.
Often these systems are subject to different types of concurrent, and
sometimes competing, constraints and objectives, such as the
availability of energy and resources, or the overall efficiency of the
resulting structure. It is therefore reasonable to assume that the
systems that we observe today are the result of a delicate balance
between contrasting forces, which can be modeled by means of an
underlying optimization process under a set of
constraints~\cite{Cancho_Sole_2003,Gastner_Newman2006,Flammini_Barthelemy_2006,Barthelemy_Flammini_2008,Barthelemy_2013}.
For instance, the emergence of scale-free networks can be explained by
simple optimization
mechanisms~\cite{Fabrikant_2002,DSouza_2007,Valverde_Sole_2002,Boguna_2012,Barabasi_2012},
while it has been found that many of the properties of biological
networks result from the simultaneous optimization of several
concurrent cost
functions~\cite{G_Nicosia_2006,Bullmore2009,Nicosia2014_worm,Seoane_2015,Seoane_Sole_2015,Shoval_Alon_2012,Sheftel_Alon_2013,Sporns_2013,Sporns_2014,G_Nicosia_2015}.
However, multi-objective optimization has not yet been linked 
to the most recent advances in network science, based on
multilayer network representations of real-world
systems~\cite{Kivela_Porter_2014,Boccaletti_2014,DeDomenico2013,Battiston2017}.
Recent studies have shown that the presence of many intertwined layers in
a network is responsible for the emergence of novel physical phenomena 
including abrupt cascading 
failures~\cite{Buldyrev2010,Radicchi2013,DSouza_2015},
superdiffusion~\cite{Gomez2013}, explosive 
synchronisation~\cite{Nicosia2017}, and the appearance of new dynamical
phases in opinion formation~\cite{Diakonova2016,Battiston2017_Axelrod}
and in epidemic processes~\cite{Sanz2014,Gleeson2016}.
Moreover, multiplexity can have an impact on practical
problems such as air traffic  
management~\cite{Lacasa_Zanin_2009,Ramasco_2013,Mantegna_Zanin_2015}
and epidemic containment~\cite{Colizza_2006,Gomes_Vespignani_2014}. As
a result, understanding how multi-layer networks
evolve~\cite{Nicosia2013,Kim2013,NicosiaNonlinear2014} 
is becoming of central importance in various 
fields. 

In this Letter we propose a model of multilayer
network growth in which the formation of links at each layer is the
result of a local Multi-Objective Optimization (MOO), i.e., a process
where two or more objective (cost) functions, often in conflict to
each other, have to be simultaneously minimized or maximized. Within
this framework, the concept of Pareto optimality naturally arises. By
introducing the dominance strict partial order~\cite{Miettinen_1999},
the solution of a MOO problem consists of a set of
\emph{non-dominated} or \emph{Pareto-optimal} points in the solution
space. Intuitively, these points represent those solutions for which
no improvement can be achieved in one objective function without
hindering the other objective functions. The collection of
non-dominated points constitutes the \emph{Pareto surface} or
\emph{Pareto front} (PF)~\cite{KDeb_2001,MOOP_2005}.

\begin{figure}[!ht]
	\includegraphics[width=3.5in]{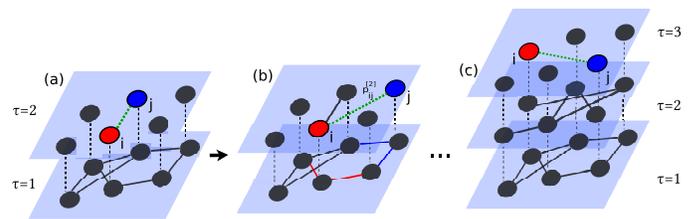}
	\caption{(Color online) Illustration of the airline growth
    model. (a) At time $\tau=2$ a new layer arrives with $K^{[2]}$
    links to be placed, and the first edge is placed uniformly at
    random among all possible pairs of nodes. (b) The remaining
    $K^{[2]}-1$ links are placed according to the probability
    $p_{ij}\lay{2}$ in Eq.~(\ref{eq:probability}). (c) The same
    procedure is repeated for each layer $\tau$ until a multiplex with
    $M$ layers is obtained.}
	\label{fig:fig1}
\end{figure}

\begin{figure*}[!t]
	\includegraphics[width=1\linewidth]{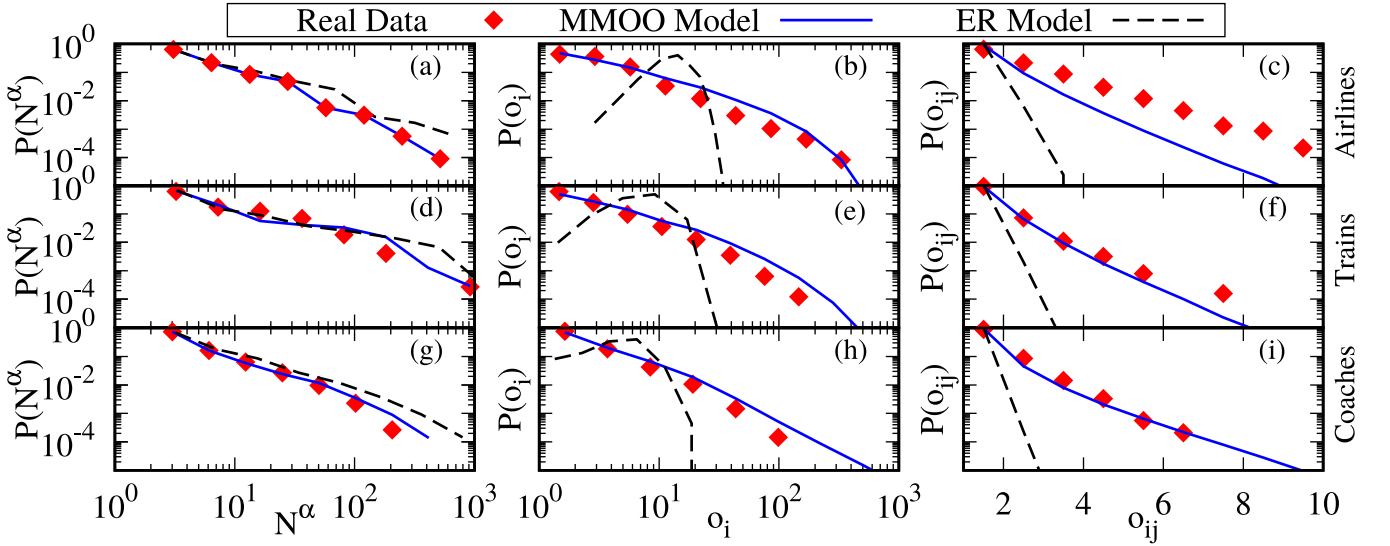}
	\caption{(Color online) Distributions of layer activity
      $N\lay{\alpha}$ (left column), node total degree $o_{i}$ (middle
      column) and edge overlap $o_{ij}$ (right column). The multiplex
      networks (red diamonds) of North America airlines (top row), UK
      train services (middle row) and UK coach services are compared
      to the corresponding multiplex networks generated by the MMOO
      model (solid blue lines) and to multiplex networks whose layers
      are \Erdos-R\'enyi graphs~\cite{Erdos} (dashed lines). The
      results shown are averaged over $10^3$ realizations (standard
      deviations are indistinguishable from the symbols).}
	\label{fig:model_distributions}
\end{figure*}

The Multiplex Multi-Objective Optimization (MMOO) model we propose is
inspired by the observation that the formation of edges in many
real-world transportation
networks~\cite{Barrat_Vespignani_2005,Guimera_Amaral_2005} is often
subject to concurrent spatial and economical
constraints~\cite{Zanin_Lillo_2013,Verma_Herrman_2014,Cook_2016}. On
the one hand there is the tendency to accumulate edges around nodes
that are already well-connected, in order to exploit the economy of
scale associated to hubs.  On the other hand, each service provider
usually tends to minimize the competition with other existing service
providers. We show that, by combining these two mechanisms,
  the MMOO model is able to reproduce quite accurately the structural
  features of three large-scale multiplex transportation systems,
  namely, the UK railway network, the UK coach network, and the six
  continental air transportation
  networks~\cite{Gardenes_Zanin_2013,Cardillo_Boccaletti_2013,Nicosia_Latora_2015,suppl_mat}. The
  MMOO model provides a reasonable explanation for the emergence of
highly-optimized heterogeneous multiplex networks.

\begin{figure*}[!t]
 	\includegraphics[width=1\linewidth]{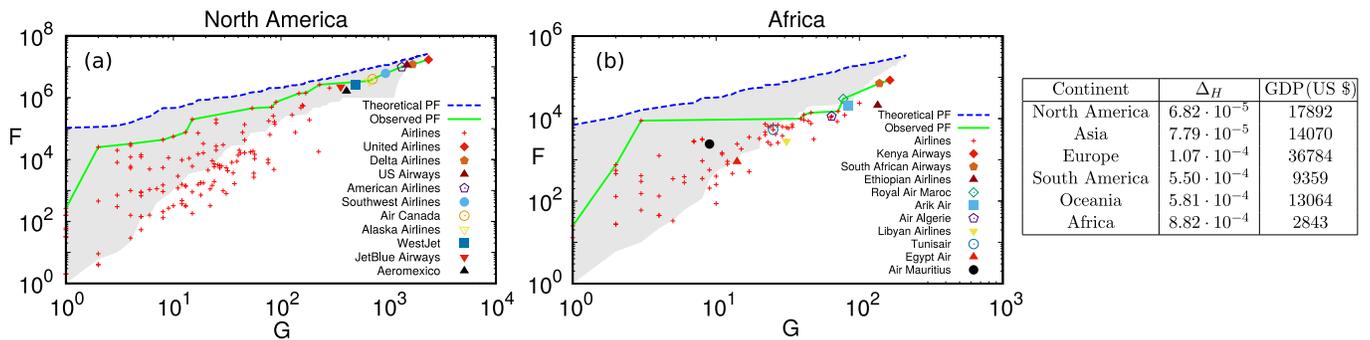}
 	\caption{(Color online) Observed (solid
      green line) and  theoretical (dashed blue line) Pareto fronts  
      for the continental airline networks of (a) North America and
      (b) Africa. The top ten airlines by number of passengers in 2013
      are highlighted. For each system, the theoretical Pareto
      front was obtained as the non-dominated points of $10^5$
      realizations of the MMOO model (the range of variability of the
      simulations is indicated by the shaded gray region). The
      potential improvement attainable by a system in the $F-G$ plane
      is measured by the normalized relative hypervolume $\Delta_H$,
      where smaller values of $\Delta_H$ correspond to more optimized
      networks.  The ranking of continents by $\Delta_H$ is reported
      in the table together with the economical performance of each continent, 
      as measured by the average GPD per capita.}
 	\label{fig:Pareto_fronts_NorthAmerica_with_tables}
\end{figure*}

\textit{Multiplex Multi-Objective Optimisation Model (MMOO). --} Let
us consider a multiplex transportation network with $N$ nodes and $M$
layers, where nodes represent locations and layers represent service
providers, e.g. airline, train or bus companies. Each layer is the
graph of routes operated by one of the service providers.  The network
can be described by a set of adjacency matrices $\{A^{[1]}, A^{[2]},
\ldots, A^{[M]}\} \in \mathbb{R}^{N\times N \times M}$, where the
entry $a_{ij}^{[\tau]}$ is equal to $1$ if $i$ and $j$ are connected
by a link at layer $\tau$ (meaning that provider $\tau$, with
$\tau=1,2,\ldots M$ operates a route between location $i$ and $j$),
while $a_{ij}^{[\tau]}=0$ otherwise. We denote by $k_i^{[\tau]}=
\sum_{j} a_{ij}^{[\tau]}$ the degree of a node $i$ at the layer
$\tau$, and by $K^{[\tau]} = \frac{1}{2}\sum_{i} k_i^{[\tau]}$ the
total number of links of layer $\tau$. An important multiplex property
of a node $i$ is the overlapping (or total) degree $o_i =\sum_\tau
k_i^{[\tau]} = \sum_j o_{ij}$, namely the total number of edges
incident on node $i$ at any of the layers of the
multiplex~\cite{Structural_Battiston_2014}, where $o_{ij}=\sum_\tau
a_{ij}^{[\tau]}$ is the overlap of edge $(i,j)$~\cite{Bianconi2013,
  Structural_Battiston_2014}, that is the number of layers at which
$i$ and $j$ are connected by an edge.

In the model we assume that service providers join the system one
after the other, each one with a predetermined number of routes that
they can operate. This means that the multi-layer network acquires a
new layer at each (discrete) time step $\tau$. When the layer joins
the system, the new provider tries to place its routes in order to
maximize its profit. To this end, a provider would prefer to have
access to as many potential customers as possible (i.e., to connect
locations with large population), whilst minimizing the competition
with other providers (i.e., to avoid to operate a route if it is
already operated by other providers). In order to mimic these two
competing drives, we set the probability to create an edge between
node $i$ and node $j$ at the new layer $\tau$ as:
 \begin{equation}
 p_{ij}^{[\tau]} \propto \frac{o_i^{[\tau-1]} o_j^{[\tau-1]} +
   c_1}{o_{ij}^{[\tau-1]}+c_2} \qquad \tau=2,\ldots,M
 \label{eq:probability}
 \end{equation}
 where $o_i^{[\tau-1]}$ and $o_{ij}^{[\tau-1]}$ are respectively the
 overlapping degree of node $i$ and the edge overlap of $(i,j)$ at
 time $\tau-1$.  The non-negative constants $c_1$ and $c_2$ allow a
 non-zero probability to create a new edge to a node that is isolated
 at all the existing layers. The rationale behind
 Eq.~(\ref{eq:probability}) is that the overlapping degree $o_i$ of
 node $i$ can be used as a proxy of the population living at that
 location. Hence, in the same spirit of the ``gravity
 model''~\cite{Gravity_model_1962,Gravity_1963}, creating a link
 between node $i$ and node $j$ with a probability proportional to the
 product $o_i o_j$ will increase the chances for a provider to access
 a large set of customers. Similarly, by requiring that
 $p_{ij}^{\lay{\tau}}$ is inversely proportional to the edge overlap
 $o_{ij}^{\lay{\tau - 1}}$ we discourage the creation of a new route
 between two locations if they are already served by a large number of
 other providers, thus modeling the tendency of providers to avoid
 competition. The two competing mechanisms we propose can be
 formalized as a MOO problem:
\begin{center}
	 	\begin{equation}
	 	\begin{cases}
	 	\displaystyle\textrm{max $\mathbf{F}$}\\ \displaystyle\textrm{min
	    $\mathbf{G}$}\\
	 	\end{cases}
	 	= 
	 	\begin{cases}
	 	\displaystyle\textrm{$F\lay{\tau}$}= \sum_{i,j:\, a_{ij}\lay{\tau}=1}(o_i \; o_j+c_1)\\
	 	\displaystyle\textrm{$G\lay{\tau}$}=\sum_{i,j:\, a_{ij}\lay{\tau}=1}(o_{ij} +c_2)\\
	 	\end{cases}
	 	\label{eq:multi-objective}
	 	\end{equation}
\end{center}
where the efficiency function $F^{[\tau]}$ accounts for the number of potential
customers, while $G^{[\tau]}$ measures the competition due to route overlaps.

The MMOO model is illustrated in Fig.~\ref{fig:fig1}. The first layer
is a connected random graph with $K\lay{1}$ edges.  At each step
$\tau$, with $\tau=2,\ldots, M$, a new layer is created.  The first of
the $K\lay{\tau}$ edges of the new layer is placed uniformly at random
among the $\binom{N}{2}$ possible edges. In order to obtain a
connected network, the remaining $K\lay{\tau}-1$ links are created
according to the probability in Eq.~(\ref{eq:probability}), yet
ensuring that one of the two endpoints of the selected edge belongs to
the connected component at that layer. The total number of links at
each of the $M$ layers are external parameters of the model.
Although considering the routes of each
company as fixed over time may look like an unrealistic oversimplification, in all the systems we have considered 
providers update their network of routes normally at a very slow rate, which justify 
our assumption to consider the routes on each layer
  as quasi-static. In fact, rearranging a set of train services or
  flights entails substantial logistic and economic investments, since
  railway licenses and airport-slots are normally allocated over time
  scales of several years.

\textit{Results. --} We have used the MMOO model to reproduce the
structure of three different multiplex transportation
  systems. The first data set includes six multiplex air
  transportation networks, each representing the airline routes
  operated in a continent. Each network has between 200
  and 1000 nodes (airports) and between 35 and 200 layers
  (carriers)~\cite{Nicosia_Latora_2015}. We have constructed the other two 
  data sets respectively from the UK national railway timetable (41
  companies operating over about 1600 stations) and from the UK
  national coach timetable (1207 companies and over 12000 coach
  stations). See Ref.~\cite{suppl_mat} for details. For each network,
we generated $10^3$ independent permutations of the sequence
$\{K\lay{1}, K\lay{2}, \ldots, K\lay{M}\}$ of the total number of
links at each layer in the data set. Then, for each permutation, we
ran $50$ independent realizations of the model.  In our simulations we
used a Metropolis-Hastings algorithm~\cite{Hastings_1970} to sample
form the distribution in Eq.~(\ref{eq:probability}).
In Fig.~\ref{fig:model_distributions} we report the distributions of layer 
activity $N\lay{\alpha}$(number
  of non-isolated nodes at each layer), total node degree $o_i$, and
  edge overlap $o_{ij}$ of the multiplex networks obtained with 
  the MMOO model, where we set $c_1 = c_2 = 1$. The two-sample
  Cramer-von-Mises statistical test~\cite{Cramer-von-Mises} provides
  convincing evidence that the synthetic distributions are compatible
  with the original ones ($p$-value $<0.01$, except for panel (c),
  where $p<0.2$). It is worth noticing that the MMOO model naturally
  reproduces the heterogeneous distribution of node total degree and
  the decreasing exponential behavior of the edge overlap $o_{ij}$,
  which respectively mirror the heavy-tailed distribution of city
  size~\cite{Zipf1949,Batty2006} and the tendency of service providers
  to reduce the competition on single routes~\cite{Cardillo_Boccaletti_2013}.
It is also possible to fine-tune the values of $c_1$ and
  $c_2$ in Eq.~(\ref{eq:probability}) in order to accurately reproduce
  also other structural properties of the three transportation
  networks, such as the distributions of node activity $B_i$ (number
  of layers at which node $i$ is not isolated) and the pattern of
  pairwise inter-layer correlations. See Ref.~\cite{suppl_mat} for
  more details.

\textit{Pareto fronts and system efficiency. -- } By considering the
multi-objective optimization framework formally defined in
Eq.~(\ref{eq:multi-objective}), it is possible to compare providers by
looking at their position in the efficiency-competition plane defined
by the two functions $F$ and $G$ and shown in
Fig.~\ref{fig:Pareto_fronts_NorthAmerica_with_tables}.  We
focused on the air transportation networks and extracted from the
empirical data the {\em observed Pareto front} of each
continental network, i.e., the set of all the non-dominated points
in the $F$-$G$ plane. Surprisingly, we found that most of the
Pareto-optimal points correspond to the most important companies in
the continent (e.g., flagship, mainline, and large low-cost
carriers), and in particular with those carrying the largest number
of passengers.

In order to quantify the potential improvement attainable by a system
in the $F$-$G$ plane, we used a multi-objective optimization
algorithm~\cite{NSGAII} to generate $10^5$ synthetic multiplex
networks for each continent. We then computed the so-called {\em
  theoretical Pareto front}, consisting of the Pareto-optimal points
resulting from all the simulations, reported in
Fig.~\ref{fig:Pareto_fronts_NorthAmerica_with_tables} as a dashed blue
line. The closer the observed PF is to the theoretical PF,
  the better the system approaches the best possible solution in the
  $F$-$G$ plane. Interestingly, the observed PF of the North American
airline network is relatively closer to its theoretical PF, while for
the African airlines we observe a larger gap between the two
curves. This means that, on average, the African airlines may obtain a
greater improvement in the $F$-$G$ plane than North American
companies. A quantitative way to associate a number to a
Pareto Front $\mathcal{P}$ is by means of the hypervolume indicator
$I_{H}(\mathcal{P})$, that is the Lebesgue measure of the union of
the rectangles defined by each point in the front $\mathcal{P}$ and
a reference point~\cite{Zitzler_1999,Zitzler_2003,suppl_mat}. The
distance between an observed PF, ${\mathcal P}^{\rm obs}$, 
and the corresponding theoretical PF, ${\mathcal P}^{\rm th}$,  
can be quantified through the relative normalized hypervolume
$\Delta_H= |I_{H_{{\mathcal P}^{\rm obs}}}- I_{H_{{\mathcal P}^{\rm 
th}}}|/(I_{H_{{\mathcal P}^{\rm th}}} K)$. By dividing
the relative hypervolume difference by the total number of routes
$K$, it is possible to compare the level of potential improvement of
two multiplex networks with respect to their corresponding
theoretical PF. We argue that the value of $\Delta_H$ can be used as
a proxy of the technological advancement of a transportation system,
with smaller values of $\Delta_H$ indicating more optimized
configurations. The Table in
Fig.~\ref{fig:Pareto_fronts_NorthAmerica_with_tables} reports the
ranking of the continents induced by $\Delta_H$, where North America and
Asia lead the pack, while Africa is lagging behind. Interestingly, that 
ranking is positively correlated with the ranking induced by continental 
GDP per capita \cite{data_GDP} (Kendall's $\tau_b=0.6$, $p \approx 9 \cdot
10^{-2}$).

\begin{figure}[!bt]
\includegraphics[width=0.9\linewidth]{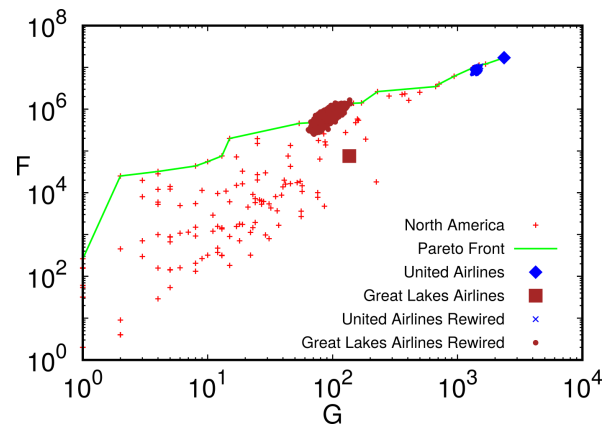}
\caption{(Color online) Reshaping the routes of two North American
  airlines in the $F$-$G$ plane. United Airlines, a Pareto-optimal company
  (blue diamonds), cannot be improved any further by our model, while
  Great Lakes Airlines (brown squares) can potentially get much
  closer to the Pareto front.}
\label{fig:rewiring_airlines}
\end{figure}

Finally, we show that our model can in principle be used by
  new companies entering the market, as a guide to place their routes
  in the most effective way. We run simulations of a slightly
  different version of the model, where all the layers are fixed and
  identical to the observed ones except for one of them, which
  represents a new service provider. The last layer is constructed
  according to Eq.~(\ref{eq:probability}). As shown in
  Fig.~\ref{fig:rewiring_airlines}, we found that in general the
  companies lying on the observed Pareto front cannot substantially
  improve their position in the $F$-$G$ plane, meaning that their
  routes have evolved over time according to an effective
  optimization process.  Conversely, our model is able to improve the
  position in the $F$-$G$ plane of sub-optimal and non-optimal
  airlines.

\textit{Conclusions. --} The introduction of multi-objective
optimization principles in the modeling of multi-layer systems allows
to obtain simple, effective explanations for the evolution of
real-world transportation networks. In particular, the systematic
exploration of the possible local improvements in the
efficiency-competition plane at the level of single carriers, that we
have used here to characterize the technological advancement
of a continent and the effectiveness of the network of single
carriers, can be employed in practice to inform the placement of new
routes, and to compare alternative expansion strategies. The proposed
methodology can be readily applied to any system whose multi-layer
structure is the result of the interactions between two or more
conflicting objective functions, paving the way to a more accurate
characterization of many different natural and man-made complex
systems.\\

V. L. acknowledges support from the EPSRC project EP/N013492/1.

\cleardoublepage
\newpage
\onecolumngrid
\section{\large{Supplementary Material}}
\normalsize
\vspace*{0.2 cm}
\twocolumngrid

\section{\label{sec:hypervolume}Hypervolume indicator}
A commonly used measure to quantitatively assess the quality of a
Pareto front in multi-objective optimization theory is the hypervolume
indicator \cite{Zitzler_1999,Zitzler_2003}, which is often referred in
literature as `size of the space covered' by the Pareto front or
`S-metric'. Given a set of Pareto optimal points $\mathcal{P} =
\{(G_1,F_1), \ldots,(G_n,F_n)\}$ in the F-G plane, the hypervolume
indicator $I_H(\mathcal{P})$ gives the area enclosed by the union of
the rectangles defined by each of the points and $(G_i,F_i)$ in
$\mathcal{P}$ and a reference point $(G_{ref},F_{ref})$. Canonical
choices for the reference point are either one of the points that
optimizes each cost function alone (also called ideal
points~\cite{Miettinen_1999,MOOP_2005}) or the worst point, i.e., the
point corresponding to the worst possible values of the two functions
to be optimized. We chose the worst point, which in our case is the
point of coordinates $(G_{max},F_{min})$.  In the paper we defined a
normalized hypervolume indicator to compare the distance of different
transportation networks with respect to their theoretical Pareto
front, which is obtained through a multi-objective optimization 
algorithm~\cite{NSGAII}. 
The metric is defined as $\Delta_H= |I_{H_{{\mathcal P}^{\rm obs}}}- 
I_{H_{{\mathcal P}^{\rm th}}}|/(I_{H_{{\mathcal P}^{\rm th}}} K)$, where 
$I_{H_{{\mathcal P}^{\rm obs}}}$ and
$I_{H_{{\mathcal P}^{\rm th}}}$ respectively represent the hypervolume 
indicator of the
observed and of the theoretical Pareto front, while $K$ is the total
number of routes in the multiplex. Notice that, since $\Delta_H$ is a
normalized relative error, the choice of the reference point has
little impact on its actual value.  We interpreted $\Delta H$ as a
measure of the potential improvement of a system towards its
theoretical Pareto front.

\begin{figure}[!ht]
	\includegraphics[width=3.2in]{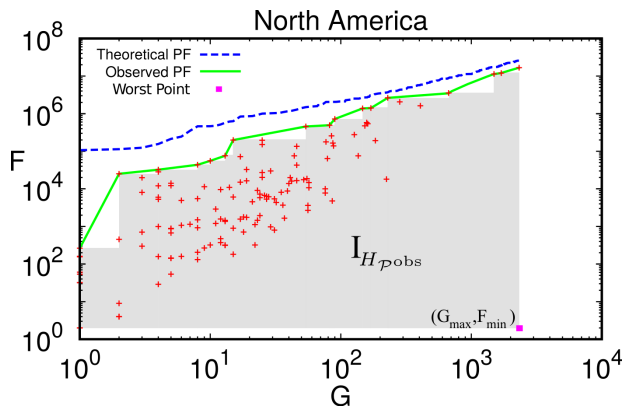}
	\caption{(Color online) Hypervolume indicator of the observed Pareto front 
		of the North 
		American airlines computed with respect to the reference point 
		$(G_{max},F_{min})$ }
\end{figure}

\section{\label{sec:dataset}Transportation network datasets}
The multiplex transportation systems we considered in the main text
are: (i) the undirected routes of the six continental airlines
(OpenFlight), (ii) the UK national railway network, and (iii) the UK
national coach network. These datasets represent the main modes of
public transportation networks. We describe below these data sets and
summarize in Table~\ref{tab:multiplexes} the basic structural
properties of the corresponding multiplex networks. Additional details
for the continental multiplex airlines can be found
in~\cite{Nicosia_Latora_2015} and the original dataset can be
downloaded at~\cite{dataset_nicosia}, while the other two datasets are
available for download at~\cite{dataset_santoro}. We considered all
these multiplexes as unweighted.

\begin{table}[!ht]
	\begin{tabular}{|c|c|c|c|}
		\hline
		Multiplex  & N & M & $\avg{N\lay{\alpha}}$\\[0.2 em] \hline
		Africa  &  238  & 84 & 9.8\\
		Asia & 795 & 213 & 24.4\\
		Europe & 593 & 175 & 21.8 \\
		North America & 1029 & 143 & 24.9\\
		Oceania & 261 & 37 & 14.1 \\
		South America & 300 & 58 &  15.1\\
		UK Coaches & 11738 & 1207 & 16.6 \\
		UK Trains & 1658 & 41 & 64.27\\
		\hline
	\end{tabular}
	\caption{The number of nodes $N$, number of layers $M$  and the average 
		layer 
		activity $\avg{N\lay{\alpha}}$ are reported for the eight multiplexes 
		analyzed in 
		this work.}
	\label{tab:multiplexes}
\end{table}

\emph{OpenFlight} - The networks of aerial routes originally used
in~\cite{Nicosia_Latora_2015} were constructed from the collaborative
free online tool OpenFlight~\cite{OpenFlights}, which allows to map
flights all around the world. For each route, the dataset contains
information about the staring airport, the destination airport, and
the company which operates each flight. Six different multiplex
networks were constructed, one for each continent (Africa, Asia,
Europe, North America, Oceania, South America), each consisting of as
many layers as airlines operating in that continent. The edges on each
layer are the direct routes operated by the corresponding airline, and
the active nodes on each layer are the airports that are connected by
at least one route operated by that company.

\emph{OpenTrainTimes} - The timetable extracted from
OpenTrainTimes~\cite{opentraintimes} contain real-time information
about the routes operated by 41 different railway companies over 2944
stations across the UK. In this case, nodes represent railway
stations, while a link exists between two nodes only if there is at
least one service between them.  The routes of OpenTrainTimes
are regularly updated using the real-time data feeds system provided
by the official network rail website~\cite{networkrail}. For each
route, we have information about the starting station, the end
station, the company which operates the service, and also the days in
which they operate their routes.  For the analysis done in the main
text, we have constructed a multiplex network with 1658 nodes (the
stations were aggregated at the city level) and 41 layers representing
the different companies.

\emph{UK coaches} - We used the UK coach data set available at
Ref.~\cite{Coaches_dataset}, which includes data from the National
Coach Services Database (NCSD). The dataset contains information about
17433 routes between 12767 locations operated by 1219 different
regional operators. In this case, nodes represent coach stations or
stops, while a link exists between two nodes only if there is at least
one connection between them. Layers represent the different regional
service providers.  Notice that for each route we have information
about the start, middle and end points, as well as the regional
operator code. Starting from this dataset we constructed a multiplex
network having 1207 regional service providers (layers) operating over
11738 stops (nodes). The analysis does not include providers operating
circular services.

\section{\label{sec:model} Details on  the MMOO model} 
In the main text, we proposed the Multiplex Multi-Objective
Optimization (MMOO) model where the probability to attach a new edge
between two nodes at a given time step is given in Eq. (1). It is
important to notice that such a probability also depends on two
non-negative constants $c_1$ and $c_2$, which allow to have a non-zero
probability when creating a new edge to a node that is isolated at all
the existing layers. We compared the synthetic and empirical
distributions for the eight transportation systems with $c_1 =1$ and
$c_2=1$. In Table~\ref{table:c_1-c_2} we report the p-values of the
corresponding two-sample Cramer-von-Mises test.

For most of the multiplex analyzed, the MMOO model with $c_1 = c_2 =
1$ performs quite well for three different structural metrics. In
particular, we find that the synthetic and empirical distributions of
layer activity $N\lay{\alpha}$ and node total degree $o_i$ are
indistinguishable. Similarly, we obtain meaningful statistical
evidence (but a lower p-value) for the distribution of edge overlap
$o_{ij}$.

To further evaluate the dependence of $c_1$ and $c_2$ on the model, we
performed extensive Monte Carlo simulations where the two constants
$c_1, c_2$ were tuned in the range $(0,10)$ with steps of $0.1$.

\begin{table}[!ht]
	\centering
	\begin{tabular}{|c|ccccc|}
		\hline
		\textbf{Multiplex} & $o_i$ & $o_{ij}$ & $N^{[\alpha]}$ & $B_i$ 
		& 
		$H_{\alpha,\beta}$ \\[0.2 em] \hline
		Africa  &  $< 0.01$ & 0.15 & $< 0.01$  & 0.2 & 0.2\\
		Asia & $< 0.01$ & 0.5 & $< 0.01$ & 0.05  & 0.05 \\
		Europe & $< 0.01$ & $< 0.01$ & $< 0.01$ & 0.05 & 0.025 \\
		North America & $< 0.01$ & 0.5 & $< 0.01$  & 0.25  & 0.75 \\
		Oceania & 0.1 & 0.15 & $< 0.01$ & 0.75  & $< 0.01$\\
		South America & $<0.01$ & 0.75 & $< 0.01$ & 0.1 & 0.25 \\
		UK Coaches & $< 0.01$ & $< 0.01$ & $< 0.01$ & 0.75 & 0.75 \\
		UK Trains & $< 0.01$ & $< 0.01$ & $ < 0.01$ & 0.5 & 0.05 \\
		\hline
	\end{tabular}
	\caption{List of the p-values obtained for the two-sample
		Cramer-von-Mises test when comparing the synthetic and empirical
		distributions of the eight transportation systems analyzed in the
		main text. Five different structural measures were tested, namely,
		the node overlap ($o_i$), the edge overlap ($o_{ij}$), the layer
		activity ($N\lay{\alpha}$), node activity ($B_i$) and the
		normalized Hamming distance ($H_{\alpha,\beta}$)
		\cite{Nicosia_Latora_2015}.  Notice that we used $c_1=1$, $c_2=1$
		in the MMOO model for obtaining the synthetic distributions.}
	\label{table:c_1-c_2}
\end{table}

\begin{figure*}[!th]
	\includegraphics[width=1 \textwidth]{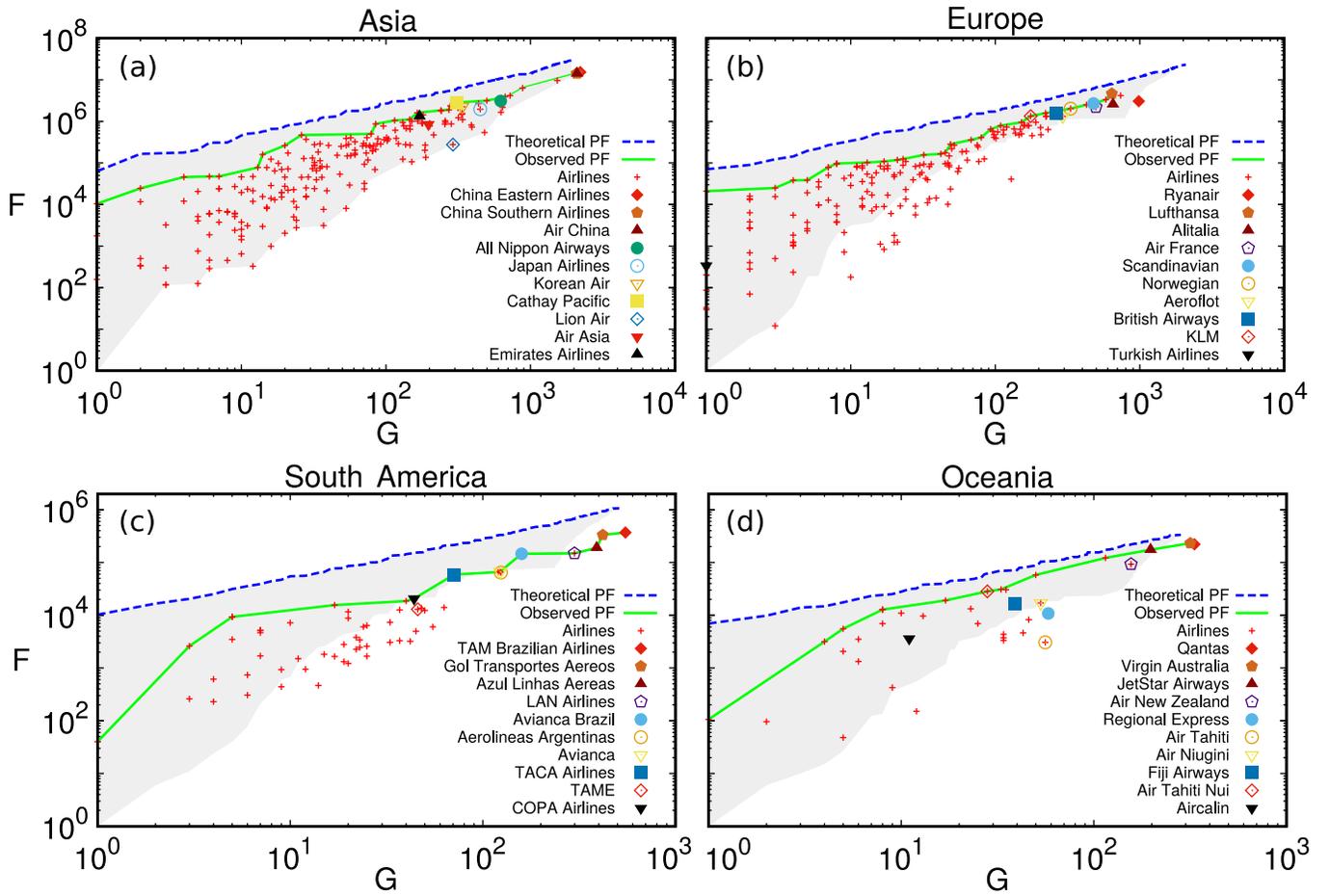}
	\caption{(Colour online) The observed Pareto front (solid green line)
		along with the top ten airlines by number of passengers for Asia
		(a), Europe (b), South America (c), and Oceania (d). For each
		continent, the Theoretical Pareto front (dashed blue line) is
		obtained by considering the non-dominated points of $10^5$
		realizations of the MMOO model (the range of variability obtained
		through the simulations is indicated by the shaded gray
		region). Notice that in almost all the cases, the top airlines of
		a continent belong to the observed Pareto front.}
	\label{fig:Pareto_continents}
\end{figure*}

In Table~\ref{table:c_1-c_2_optimised} we report, for each system, the
values of $c_1$ and $c_2$ for which the two-sample Cramer-von-Mises
test provides the best (lowest) p-values for the distributions of node
total degree $o_i$, edge overlap $o_{ij}$, layer activity
$N\lay{\alpha}$, node activity $B_i$ and pairwise inter-layer
correlation $H_{\alpha, \beta}$~(see Ref.~\cite{Nicosia_Latora_2015}
for details). For the sake of clarity, we also report in
Fig.~\ref{fig:Pareto_continents} the placement of each airline in the
efficiency-competition plane and the corresponding Pareto front for
the four continental airlines of Asia, Europe, South America and
Oceania. In each plot we highlighted the top ten companies by number
of passengers in 2012. Notice that whenever an airline group is listed
in the top ten airways, we report each element of the group.

\begin{table}[hb!]
	\centering
	\begin{tabular}{|c|ccccc|c|c|}
		\hline
		\textbf{Multiplex} & $o_i$ & $o_{ij}$ & $N^{[\alpha]}$ & $B_i$
		& $H_{\alpha,\beta}$ & $c_1$ & $c_2$\\[0.2 em] \hline
		Africa  &  $< 0.01$ & $< 0.01$ & $< 0.01$ & 0.1 & 0.25 & 0.2 & 9.3 \\
		Asia & $< 0.01$ & $<0.01$ & $< 0.01$ & $<0.01$ &  0.1 & 0.2 & 3.0 \\
		Europe & $< 0.01$ & $< 0.01$ & $<0.01$ & $< 0.01$ & 0.025 & 
		1.0 & 8.2\\
		North America & $< 0.01$ & $< 0.01$ & $< 0.01$ & 0.1 &  0.75& 
		0.1 & 9.0\\
		Oceania &  0.025 & 0.05 & $< 0.01$ & 0.75 &  $< 0.01$ & 1.8 & 6.0\\
		South America &  $< 0.01$ & $< 0.01$ & $< 0.01$ & 0.1 & 0.5 & 0.2 & 
		9.0\\
		UK Coaches & $< 0.01$ & $< 0.01$ & $< 0.01$ & 0.5  & 0.75 & 
		4.0 & 6.5\\
		UK Trains & $< 0.01$ & $< 0.01$ & $< 0.01$ & 0.25 & $< 0.01$  & 
		2.1 & 5.3\\
		\hline	
	\end{tabular}
	\label{table:c_1-c_2_optimised}
	\caption{Values of $c_1$ and $c_2$ for which we obtain the best
		(lowest) p-values of the Cramer-von-Mises test for all the five
		structural multiplex metrics.}
\end{table}



\end{document}